\documentclass[amsmath,aps,showpacs,a4paper,10pt]{revtex4}
 \usepackage{epsf}
 \usepackage{graphicx}    % Include figure files
 \usepackage{enumerate}   % enumerate itemize

\begin{document}

 \newcommand{\be}[1]{\begin{equation}\label{#1}}
 \newcommand{\ee}{\end{equation}}
 \newcommand{\bea}{\begin{eqnarray}}
 \newcommand{\eea}{\end{eqnarray}}
 \def\disp{\displaystyle}

 \def\gsim{ \lower .75ex \hbox{$\sim$} \llap{\raise .27ex \hbox{$>$}} }
 \def\lsim{ \lower .75ex \hbox{$\sim$} \llap{\raise .27ex \hbox{$<$}} }

 \begin{titlepage}

 \begin{flushright}
 arXiv:2002.10189
 \end{flushright}

 \title{\Large \bf Reconstructing the Fraction of Baryons
 in the Intergalactic Medium with Fast Radio Bursts via
 Gaussian Processes}

 \author{Da-Chun~Qiang\,}
 \email[\,email address:\ ]{875019424@qq.com}
 \affiliation{School of Physics,
 Beijing Institute of Technology, Beijing 100081, China}

 \author{Hao~Wei\,}
 \email[\,Corresponding author;\ email address:\ ]{haowei@bit.edu.cn}
 \affiliation{School of Physics,
 Beijing Institute of Technology, Beijing 100081, China}

 \begin{abstract}\vspace{1cm}
 \centerline{\bf ABSTRACT}\vspace{2mm}
 Fast radio bursts (FRBs) are a promising new probe for astronomy and
 cosmology. Thanks to their extragalactic and cosmological origin, FRBs
 could be used to study the intergalactic medium (IGM) and the cosmic
 expansion. It is expected that numerous FRBs with identified
 redshifts will be available in the near future through the identification
 of their host galaxies or counterparts. $\rm DM_{IGM}$, the contribution
 from IGM to the observed dispersion measure (DM) of FRB, carries the
 key information about IGM and the cosmic expansion history. We can
 thus study the evolution of the universe by using FRBs with identified
 redshifts. In the present work, we are interested in the fraction of
 baryon mass in the IGM, $f_{\rm IGM}$, which is useful to study the cosmic
 expansion and the problem of the ``\,missing baryons\,''. We propose to
 reconstruct the evolution of $f_{\rm IGM}$ as a function of redshift
 $z$ with FRBs via a completely model-independent method, namely Gaussian
 processes. Since there is not a large sample of FRBs with identified
 redshifts, we use simulated FRBs instead. Through various simulations,
 we show that this methodology works well.
 \end{abstract}

 % https://publishing.aip.org/publishing/pacs/pacs-2010-regular-edition
 % https://ufn.ru/en/pacs/all/

 \maketitle

 \end{titlepage}

 \renewcommand{\baselinestretch}{1.0}

%============================= section 1 ===================================

\section{Introduction}\label{sec1}

Fast radio bursts (FRBs) have become a promising field in astronomy
 and cosmology~\cite{NAFRBs,Lorimer:2018rwi,Keane:2018jqo,
 Kulkarni:2018ola,Burke-Spolaor:2018xoa,Pen:2018ilo,Macquart:2018fhn,
 Caleb:2018ygr} since their discovery just over a decade
 ago~\cite{Lorimer:2007qn}. The key measured quantity of FRBs is the
 dispersion measure (DM). The large DMs of observed FRBs well in excess
 of the Galactic value strongly suggested a cosmological
 origin~\cite{Dolag:2014bca} as is now known to be the case through the
 localization of a handful of FRBs to host galaxies~\cite{Tendulkar:2017vuq,
 Bannister2019,Ravi:2019alc,Marcote:2020ljw}. As a crude rule
 of thumb, the redshift of FRB $z\sim {\rm DM}/(1000\;\rm pc\hspace{0.24em}
 cm^{-3})$~\cite{Lorimer:2018rwi}. Currently, the DMs of the observed FRBs
 are in the range $100\sim 2600\;\rm pc\hspace{0.24em}cm^{-3}$ approximately
 \cite{Petroff:2016tcr}, and hence one can infer their redshifts in the
 approximate redshift range $0.1\lesssim z\lesssim 2.6$. There are several
 possibilities to identify the redshifts of FRBs. For repeating FRBs,
 precise localizations have been made to host galaxies. The redshift of the
 first known repeating FRB (namely FRB 121102~\cite{Spitler:2016dmz,
 Marcote:2017wan,Chatterjee:2017dqg,Tendulkar:2017vuq}) has been identified
 as $z=0.19273$~\cite{Tendulkar:2017vuq}. More and more repeating FRBs have
 been found, such as the other 18 repeating FRBs reported by CHIME/FRB
 Collaboration~\cite{Amiri:2019bjk,Andersen:2019yex,Fonseca:2020cdd}. On the
 other hand, the redshifts of FRBs can also be precisely determined if their
 afterglows or counterparts (e.g. gamma-ray bursts (GRBs) or gravitational
 wave events (GWs)) are observed, although no FRB has been seen in any other
 band than radio to date. Very recently, precise localizations of host
 galaxies of FRBs have been obtained even for the non-repeating FRBs, such
 as FRB 180924 which has been localized to a massive galaxy at redshift
 $z=0.3214$~\cite{Bannister2019} using ASKAP. Another non-repeating
 FRB 190523 has been localized to a few-arcsecond region containing a single
 massive galaxy at redshift $z=0.66$~\cite{Ravi:2019alc} using DSA-10.
 Currently, several projects designed to detect and localize FRBs with
 arcsecond accuracy in real time are under construction/proposition, for
 example, DSA-10~\cite{DSA-10}, DSA-2000~\cite{DSA-2000},
 UTMOST-2D~\cite{UTMOST}, MeerKAT~\cite{Booth:2009ex,Johnston:2020qxo,
 MeerTRAP}, and LOFAR~\cite{vanHaarlem:2013dsa}. It is expected that
 numerous FRBs with identified redshifts will be available in the future.
 Since they are at cosmological distances, it is justified and
 well-motivated to study cosmology by using FRBs.

For a cold plasma~\cite{Rybicki:1979} (see also
 e.g.~\cite{Deng:2013aga,Yang:2016zbm,Ioka:2003fr,Inoue:2003ga,Qiang:2019zrs}),
 an electromagnetic signal of frequency $\nu$ propagates through an ionized
 medium (plasma) with a velocity less than the speed of light in vacuum~$c$,
 and hence this signal with frequency $\nu\gg\nu_p$ is delayed relative to
 a signal in vacuum, where $\nu_p$ is the plasma frequency. In practice, it
 is convenient to measure the time delay between two frequencies $\nu_1$
 and $\nu_2$, which is given by~\cite{Deng:2013aga,Yang:2016zbm,Ioka:2003fr,
 Inoue:2003ga,Qiang:2019zrs}
 \be{eq1}
 \Delta t=\frac{e^2}{2\pi m_{e\,}c}\left(
 \frac{1}{\nu_{1}^2}-\frac{1}{\nu_{2}^2}\right)
 \int\frac{n_{e,\,z}}{1+z}\,dl\equiv
 \frac{e^2}{2\pi m_{e\,}c}\left(
 \frac{1}{\nu_{1}^2}-\frac{1}{\nu_{2}^2}\right){\rm DM}\,,
 \ee
 where $n_{e,\,z}$ is the number density of free electrons in the medium
 (given in units of $\rm cm^{-3}$) at redshift $z$, $m_e$ and $e$ are the
 mass and charge of electron, respectively. Using Eq.~(\ref{eq1}), one can
 get the column density of the free electrons ${\rm DM}\equiv\int n_{e,\,z}
 /(1+z)\,dl$ by measuring the time delay $\Delta t$ between two frequencies
 $\nu_1$ and $\nu_2$. It is worth noting that the distance $dl$ along the
 path in DM records the expansion history of the universe. Thus, the
 dispersion measure DM plays a key role in the FRB cosmology.

The observed DM of FRB can be separated into three components
 \cite{Deng:2013aga,Yang:2016zbm,Qiang:2019zrs,
 Gao:2014iva,Zhou:2014yta,Yang:2017bls,Li:2019klc,Wei:2019uhh}
 \be{eq2}
 \rm DM_{obs}=DM_{MW}+DM_{IGM}+DM_{HG}\,,
 \ee
 where $\rm DM_{MW}$, $\rm DM_{IGM}$, and $\rm DM_{HG}$ are the
 contributions from the Milky Way, the intergalactic medium (IGM), and the
 host galaxy (HG, including interstellar medium of HG and the near-source
 plasma) of the FRB, respectively. In particular, $\rm DM_{MW}$ can be well
 constrained with pulsar data~\cite{Taylor:1993my,Manchester:2004bp}. For a
 well-localized FRB, the corresponding $\rm DM_{MW}$ can be estimated with
 reasonable certainty~\cite{Cordes:2002wz,Cordes:2003ik,YMW16}. Thus, it is
 convenient to introduce the extragalactic DM of an FRB as the observed
 quantity~\cite{Yang:2016zbm,Yang:2017bls,Qiang:2019zrs,Li:2019klc},
 \be{eq3}
 \rm DM_E\equiv DM_{obs}-DM_{MW}=DM_{IGM}+DM_{HG}\,,
 \ee
 by subtracting this ``\,known\,'' $\rm DM_{MW}$ from $\rm DM_{obs}$ and
 using Eq.~(\ref{eq2}). The main contribution to DM of FRB comes from
 IGM. As is shown in e.g.~\cite{Deng:2013aga,Yang:2016zbm,
 Qiang:2019zrs,Li:2019klc,Wei:2019uhh}, the mean of $\rm DM_{IGM}$ is
 given by
 \be{eq4}
 \langle{\rm DM_{IGM}}\rangle=\frac{3cH_0\Omega_{b,\,0}}{8\pi G m_p}
 \int_0^z\frac{f_{\rm IGM}(\tilde{z})\,f_e(\tilde{z})\left(1+
 \tilde{z}\right)d\tilde{z}}{E(\tilde{z})}\,,
 \ee
 where $\Omega_{b,\,0}= 8\pi G\rho_{b,\,0}/(3H_0^2)$ is the present
 fractional density of baryons (the subscript ``\,0\,'' indicates the
 present value of the corresponding quantity), $H_0$ is the Hubble constant,
 $m_p$ is the mass of proton, $E\equiv H/H_0$ (in which $H\equiv\dot{a}/a$
 is the Hubble parameter, $a=(1+z)^{-1}$ is the scale factor, a dot denotes
 the derivative with respect to cosmic time $t$), $f_{\rm IGM}$
 is the fraction of baryon mass in IGM, and
 \be{eq5}
 f_e\equiv Y_{\rm H}\,\chi_{e,\,\rm H}(z)+
 \frac{1}{2}\,Y_{\rm He}\,\chi_{e,\,\rm He}(z)\,,
 \ee
 in which the hydrogen (H) mass fraction $Y_{\rm H}=(3/4)\,y_1$, and the
 helium (He) mass fraction $Y_{\rm He}=(1/4)\,y_2$, where $y_1\sim 1$
 and $y_2\simeq 4-3y_1\sim 1$ are the hydrogen and helium mass fractions
 normalized to the typical values $3/4$ and $1/4$, respectively. Their
 ionization fractions $\chi_{e,\,\rm H}(z)$ and $\chi_{e,\,\rm He}(z)$ are
 both functions of redshift~$z$. It is expected that intergalactic hydrogen
 and helium are fully ionized at redshifts $z\lesssim 6$ and $z\lesssim 3$
 \cite{Meiksin:2007rz,Becker:2010cu} (see also e.g.~\cite{Shull:2010ku}),
 respectively. So, for FRBs at redshifts $z\leq 3$, intergalactic hydrogen
 and helium are both fully ionized, and hence $\chi_{e,\,\rm H}(z)=
 \chi_{e,\,\rm He}(z)=1$. In this case, $f_e(z)\simeq 7/8$, and
 then Eq.~(\ref{eq4}) becomes
 \be{eq6}
 \langle{\rm DM_{IGM}}\rangle=Q_{\rm IGM}\int_0^z
 \frac{f_{\rm IGM}(\tilde{z})\left(1+\tilde{z}
 \right)d\tilde{z}}{E(\tilde{z})}\,,
 \ee
 where
 \be{eq7}
 Q_{\rm IGM}\equiv\frac{3cH_0\Omega_{b,\,0}f_e}{8\pi G m_p}\,.
 \ee
 Note that $\rm DM_{IGM}$ will deviate from
 $\langle{\rm DM_{IGM}}\rangle$ if the plasma density fluctuations are
 taken into account~\cite{McQuinn:2013tmc} (see
 also e.g.~\cite{Ioka:2003fr,Jaroszynski:2018vgh}). On the other hand,
 the contribution from the host galaxy of the FRB, i.e. $\rm DM_{HG}$,
 is poorly known. For an FRB at redshift $z$, its observed $\rm DM_{HG}$
 should be redshifted (see e.g.~\cite{Yang:2016zbm,Gao:2014iva,
 Yang:2017bls,Zhou:2014yta,Qiang:2019zrs,Li:2019klc}), namely
 \be{eq8}
 {\rm DM_{HG}}={\rm DM_{HG,\,loc}}/(1+z)\,,
 \ee
 where $\rm DM_{HG,\,loc}$ is the local DM of FRB host galaxy. In the
 literature (e.g.~\cite{Yang:2017bls,Qiang:2019zrs}), the local DM of FRB
 host galaxy might be assumed to have no significant evolution with
 redshift, namely $\rm DM_{HG,\,loc}$ is a constant independent
 of redshift $z$.

Clearly, the fraction of baryons in IGM (namely $f_{\rm IGM}$) and the
 local value of $\rm DM_{HG}$ (namely $\rm DM_{HG,\,loc}$) will play the
 key roles when we use the observed $\rm DM_E$ to study cosmology. However,
 they are both poorly known. It is of interest to obtain them from the
 observational data. Furthermore, studies of $f_{\rm IGM}$
 are also important to the problem of ``\,missing baryons\,''
 (see e.g.~\cite{Macquart:2018fhn,McQuinn:2013tmc,Cen:1998hc,
 Bregman:2007ac,Shull:2011aa}). Until very recently, censuses of the nearby
 universe fail to account for roughly half of the entire baryonic matter
 content that is estimated to exist on the basis of both cosmological
 theory and measurements of the hydrogen density in intergalactic gas
 10 billion years ago~\cite{Macquart:2018fhn,McQuinn:2013tmc,Cen:1998hc,
 Bregman:2007ac,Shull:2011aa}. In contrast to the other observables,
 every diffuse ionized baryon along a sightline contributes equally to
 DM~\cite{Macquart:2018fhn,McQuinn:2013tmc}. Thus, the constraints on
 the fraction of baryons in IGM (namely $f_{\rm IGM}$) by using FRBs are
 unique and helpful to address this missing baryons problem.

In the literature (e.g.~\cite{Yang:2016zbm,Yang:2017bls,Gao:2014iva,
 Qiang:2019zrs}), a redshift independent $f_{\rm IGM}$ (say, 0.83) is
 usually assumed. However, in principle $f_{\rm IGM}$ should be a function
 of redshift $z$. It is of interest to consider the evolution
 of $f_{\rm IGM}(z)$. In~\cite{Li:2019klc}, a linear parameterization for
 $f_{\rm IGM}(z)$ with respect to the scale factor $a$ was considered,
 namely $f_{\rm IGM}(z)=f_{\rm IGM,\,0}\,(1+\alpha\,(1-a))=
 f_{\rm IGM,\,0}\,(1+\alpha\,z/(1+z))$. In~\cite{Wei:2019uhh},
 $f_{\rm IGM}(z)$ divided into five redshift bins was considered. We note
 that in the first case~\cite{Li:2019klc} a specific function form for
 $f_{\rm IGM}(z)$ is assumed {\it a~prior} and hence it is not
 so model-independent in fact, while in the second case~\cite{Wei:2019uhh}
 the binned $f_{\rm IGM}(z)$ is not a continuous function of redshift $z$
 and hence cannot reconstruct the smooth evolution of $f_{\rm IGM}(z)$.
 In the present work, we propose a completely model-independent method
 to reconstruct $f_{\rm IGM}(z)$. As is well known, by using Gaussian
 processes~\cite{Rasmussen:2006,GPweb,Seikel:2012uu,Seikel:2013fda,
 Yin:2018mvu,Li:2019nux,Cai:2019bdh,Lin:2019cuy,Jesus:2019nnk,
 Belgacem:2019zzu,Zhang:2018gjb}, the goal function can be reconstructed
 directly from the input data without assuming a particular function form
 or parameterization. Derivatives of the function can also be reliably
 reconstructed. Obviously, this is indeed model-independent. Here, we
 try to reconstruct the evolution of $f_{\rm IGM}(z)$ with FRBs via
 Gaussian processes.

The rest of this paper is organized as follows. In Sec.~\ref{sec2}, we
 describe the methodology to reconstruct the evolution of $f_{\rm IGM}(z)$,
 and briefly introduce the key points of Gaussian processes. In
 Sec.~\ref{sec3}, we test this new method by reconstructing $f_{\rm IGM}(z)$
 with the simulated FRBs and the observational Pantheon sample of type Ia
 supernovae (SNIa). In Sec.~\ref{sec4}, some brief concluding remarks
 are given.

%============================= section 2 ===================================

\section{Methodology to reconstruct the evolution
 of $f_{\rm IGM}(z)$}\label{sec2}

%=========================== section 2.1 =================================

\subsection{Formalism}\label{sec2a}

Initially, we attempt to find a formalism to reconstruct $f_{\rm IGM}(z)$.
 Obviously, $f_{\rm IGM}(z)$ enters into DM through $\rm DM_{IGM}$.
 Differentiating Eq.~(\ref{eq6}), we obtain
 \be{eq9}
 \langle{\rm DM_{IGM}}\rangle^\prime=Q_{\rm IGM}
 \,\frac{f_{\rm IGM}(z)\left(1+z\right)}{E(z)}\,,
 \ee
 where a prime denotes the derivative with respect to redshift $z$. From
 Eqs.~(\ref{eq3}) and (\ref{eq8}), we have
 \be{eq10}
 \langle{\rm DM_E}\rangle\left(1+z\right)=\langle{\rm DM_{IGM}}\rangle
 \left(1+z\right)+\langle{\rm DM_{HG,\,loc}}\rangle\,.
 \ee
 Differentiating Eq.~(\ref{eq10}), we find that
 \be{eq11}
 \left[\langle{\rm DM_E}\rangle\left(1+z\right)\,\right]^{\,\prime}
 =\langle{\rm DM_E}\rangle^\prime\left(1+z\right)+\langle{\rm DM_E}\rangle
 =\langle{\rm DM_{IGM}}\rangle^\prime\left(1+z\right)+
 \langle{\rm DM_{IGM}}\rangle\,.
 \ee
 Substituting Eq.~(\ref{eq9}) into Eq.~(\ref{eq11}) and using
 Eq.~(\ref{eq10}), it is easy to see that
 \be{eq12}
 Q_{\rm IGM}\,\frac{f_{\rm IGM}(z)\left(1+z\right)^3}{E(z)}=
 \langle{\rm DM_E}\rangle^\prime
 \left(1+z\right)^2+\langle{\rm DM_{HG,\,loc}}\rangle\,.
 \ee
 Further, noting $\left.{\rm DM_{IGM}}\right|_{z=0}=0$ by
 definition, from Eq.~(\ref{eq10}), we have
 \be{eq13}
 \langle{\rm DM_{HG,\,loc}}\rangle=
 \left.\langle{\rm DM_E}\rangle\right|_{z=0}\,.
 \ee
 Thus, once $\langle{\rm DM_E}\rangle(z)$,
 $\langle{\rm DM_E}\rangle^\prime (z)$ and $E(z)$ have been reconstructed,
 $f_{\rm IGM}(z)$ and $\langle{\rm DM_{HG,\,loc}}\rangle$ are at hand.
 However, on the observational side, we only have the observed
 $\rm DM_E$ rather than $\langle{\rm DM_E}\rangle$. In this case, we
 instead reconstruct $f_{\rm IGM}(z)$ by using
 \be{eq14}
 f_{\rm IGM}(z)=\frac{E(z)}{\rm Q_{IGM}}\left(1+z\right)^{-3}\left[\,
 {\rm DM_E^{\,\prime}}\left(1+z\right)^2+{\rm DM_{HG,\,loc}}\right]\,,
 \ee
 in which
 \be{eq15}
 {\rm DM_{HG,\,loc}}= \left. {\rm DM_E}\right|_{z=0}\,.
 \ee
 We can reconstruct $\rm DM_E$ and $\rm DM_E^{\,\prime}$ as functions of
 redshift $z$ from the observed $\rm DM_E$ data of FRBs by using Gaussian
 processes, and then obtain $\rm DM_{HG,\,loc}$ from the reconstructed
 ${\rm DM_E}(z)$ at $z=0$. On the other hand, we can also reconstruct
 $E(z)$ from the observational data of SNIa by using Gaussian processes. The
 luminosity distances of SNIa are given by $d_L(z_{\rm cmb},\,z_{\rm hel})=
 \left(c/H_0\right)\left(1+z_{\rm hel}\right)D(z_{\rm cmb})$
 (see e.g.~\cite{Conley:2011ku,Wang:2015tua,Li:2016dqg,Deng:2018jrp,
 Deng:2018yhb}), where $z_{\rm cmb}$ and $z_{\rm hel}$ are the CMB restframe
 redshift and the heliocentric redshift of SNIa, respectively. Note that we
 consider a flat Friedmann-Robertson-Walker (FRW) universe throughout. In
 this case, $D(z)=\int_0^z d\tilde{z}/E(\tilde{z})$, and hence
 $E=1/D^\prime$. Finally, using Eq.~(\ref{eq14}), we can reconstruct
 $f_{\rm IGM}(z)$ from the observational data of FRBs and SNIa via Gaussian
 processes.

%=========================== section 2.2 =================================

\subsection{The key points of Gaussian processes}\label{sec2b}

Gaussian processes~\cite{Rasmussen:2006,GPweb,Seikel:2012uu,Seikel:2013fda}
 can provide an algorithm for machine learning. By using Gaussian processes,
 the goal function can be reconstructed directly from the input data without
 assuming a particular function form or parameterization. Derivatives
 of the goal function can also be reconstructed reliably. Following
 e.g.~\cite{Rasmussen:2006,Seikel:2012uu,Seikel:2013fda}, here we briefly
 introduce the key points of Gaussian processes. A Gaussian process is the
 generalization of a Gaussian distribution. While the latter is the
 distribution of a random variable, Gaussian process describes
 a distribution over functions. At each point $z$, the reconstructed
 function $f(z)$ is described by a Gaussian distribution. Function values at
 different points $z$ and $\tilde{z}$ are not independent of each other,
 but are related by a covariance function (also called the kernel function
 in the literature) $k(z,\,\tilde{z})$, which depends on the hyperparameters
 such as $\sigma_{\! f}$ and $\ell$. The observational data can also be
 described by a Gaussian process, assuming the errors are Gaussian. For a
 given covariance function and hyperparameters, the reconstructed function
 is determined by the covariances between the observational data and the
 points $\{z_i\}$ at which the function $f(z)$ will be reconstructed.
 Note that in Gaussian processes, the hyperparameters are determined
 (trained) by the observational data (this can be done by maximizing the
 marginal likelihood or marginalizing over the hyperparameters). In
 addition, the derivatives $f^{\,\prime}(z)$, $f^{\,\prime\prime}(z)$,
 $f^{\,\prime\prime\prime}(z)$ ... can also be reconstructed by performing
 Monte Carlo samplings from a multivariate Gaussian distribution. We refer
 to e.g.~\cite{Rasmussen:2006,Seikel:2012uu,Seikel:2013fda} for
 technical details.

In this work, we implement Gaussian processes by using the publicly
 available code GaPP (Gaussian Processes in Python)~\cite{Seikel:2012uu}.
 In Gaussian processes, there exist many options for the covariance
 function $k(z,\,\tilde{z})$. In practice, the choices of covariance
 function only make fairly small difference (see e.g.~\cite{Seikel:2013fda,
 Yin:2018mvu}). So, in this work we choose to use the simplest one (which
 is also the most popular choice in the literature), namely the squared
 exponential (or, Gaussian) covariance function
 (see e.g.~\cite{Rasmussen:2006,Seikel:2012uu,Seikel:2013fda})
 \be{eq16}
 k(z,\,\tilde{z})=\sigma_{\! f}^2\exp\left(-\frac{(z-\tilde{z})^2\,}
 {2\,\ell^2}\right)\,.
 \ee

%============================= Fig. 1 =================================

 \begin{center}
 \begin{figure}[tb]
 \centering
 \vspace{-9mm}  % used here just for a comfortable typesetting
 \includegraphics[width=0.85\textwidth]{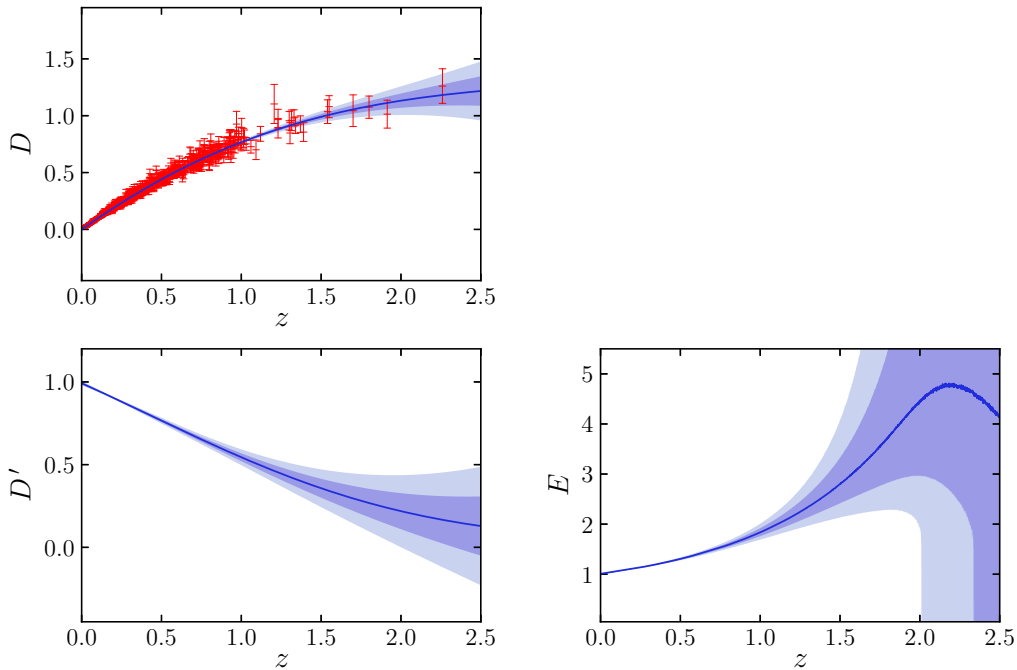}
 \caption{\label{fig1} The reconstructed $D=d_L\left(H_0/c\right)/\left(1+
 z\right)\,$, $D^{\,\prime}$, and $E=1/D^{\,\prime}$ as functions of
 redshift $z$ from the observational Pantheon sample consisting of 1048 SNIa
 via Gaussian processes. The mean and $1\sigma$, $2\sigma$ uncertainties
 are indicated by the blue solid lines and the shaded regions,
 respectively. The observational $D_{\rm obs}$ data with red error bars
 are also plotted in the top panel. See Sec.~\ref{sec3a} for details.}
 \end{figure}
 \end{center}

%======================================================================

 \vspace{-9mm}  % used here just for a comfortable typesetting

%============================= section 3 ===================================

\section{Reconstructing $f_{\rm IGM}(z)$ with the simulated
 FRBs}\label{sec3}

%=========================== section 3.1 =================================

\subsection{The reconstructed $E(z)$ from the observational data of
 SNIa}\label{sec3a}

To obtain $f_{\rm IGM}(z)$ by using Eq.~(\ref{eq14}), we should initially
 reconstruct the cosmic expansion history characterized by $E(z)$. As is
 well known, SNIa are suitable indicators of the cosmic expansion history.
 It is thus natural to reconstruct $E(z)$ from the observational data of
 SNIa by using Gaussian processes, as stated at the end of Sec.~\ref{sec2a}.
 Following~\cite{Yin:2018mvu}, we use the observational Pantheon sample
 \cite{Scolnic:2017caz,Pantheondata,Pantheonplugin,Panupdated}
 consisting of 1048 SNIa, which is the largest spectroscopically confirmed
 SNIa sample to date. Its observational data are given in terms of the
 corrected bolometric apparent magnitude $m$. The quantity $D$ introduced at
 the end of Sec.~\ref{sec2a} is related to $m$ according to (see e.g.
 \cite{Conley:2011ku,Wang:2015tua,Li:2016dqg,Deng:2018jrp,Deng:2018yhb})
 \be{eq17}
 m(z_{\rm cmb},\,z_{\rm hel})=5\,\log_{10}\left(\left(1+z_{\rm hel}\right)
 D(z_{\rm cmb})\right)+{\cal M}\,,
 \ee
 where $\cal M$ is a nuisance parameter representing some combination of
 the absolute magnitude $M$ and $H_0$. One can convert the observational
 $m$ data given in the Pantheon plugin~\cite{Pantheonplugin,Panupdated}
 into the $D_{\rm obs}$ data, while their covariance matrices are related
 by the propagation of uncertainty~\cite{poucov}, $\boldsymbol{C}_D=
 \boldsymbol{J}\boldsymbol{C}_m \boldsymbol{J}^{\,T}$, where
 $\boldsymbol{J}$ is the Jacobian matrix. We use the full covariance
 matrix including the systematic uncertainties. It is worth noting that
 the data of the Pantheon SNIa sample have been slightly
 updated~\cite{Panupdated} at the end of 2018, and hence there might be
 minor differences between the results from the old and the updated
 Pantheon datasets. Fitting the flat $\Lambda$CDM model to the updated
 Pantheon SNIa dataset, we obtain the best-fit ${\cal M}=23.80854156$ (see
 Appendix C of~\cite{Conley:2011ku} for technical details), and then adopt
 it as a fiducial value. We can reconstruct $D(z)$ and $D^\prime(z)$ from
 the observational $D_{\rm obs}$ data via Gaussian processes, and hence
 $E=1/D^\prime$ is ready. We present them in Fig.~\ref{fig1}. In particular,
 this reconstructed $E(z)$ will be used in Eq.~(\ref{eq14}) to reconstruct
 $f_{\rm IGM}(z)$.

%============================= Fig. 2 =================================

 \begin{center}
 \begin{figure}[tb]
 \centering
 \vspace{-8mm}  % used here just for a comfortable typesetting
 \includegraphics[width=0.85\textwidth]{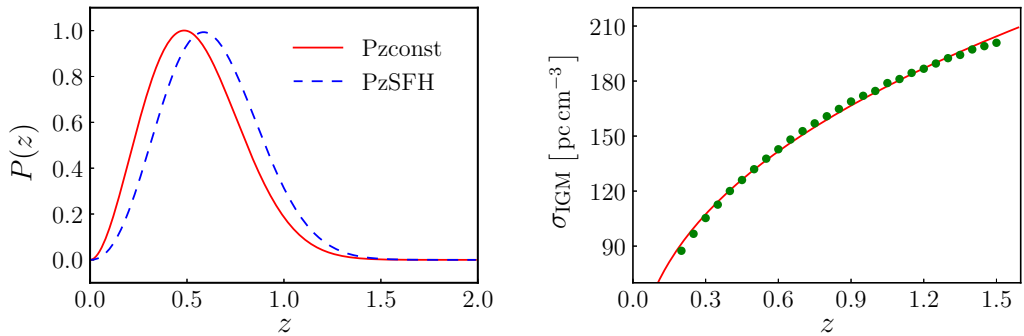}
 \caption{\label{fig2} Left panel: The redshift distributions Pzconst
 (red solid line) and PzSFH (blue dashed line), normalized with respect
 to the maximum. Right panel: $\sigma_{\rm IGM}$ versus redshift $z$.
 The 27 green dots are reproduced from the bottom panel of
 Fig.~1 of~\cite{McQuinn:2013tmc}. The red solid line is plotted according
 to Eq.~(\ref{eq22}). See Sec.~\ref{sec3b} for details.}
 \end{figure}
 \end{center}

%======================================================================

 \vspace{-12mm}  % used here just for a comfortable typesetting

%=========================== section 3.2 =================================

\subsection{Simulating FRBs}\label{sec3b}

As mentioned above, we have only a few FRBs with identified redshifts to
 date, due to the relatively small areas of sky that can be monitored and
 the need for telescope arrays in order to provide host galaxy localization.
 The lower-limit estimates for the number of FRB events are a few thousands
 per sky per day~\cite{Keane:2018jqo,Bhandari:2017qrj}. Even conservatively,
 the FRB event rate floor derived from the pre-commissioning of CHIME/FRB is
 $3\times 10^2$ events per day \cite{Amiri:2019qbv}. Several projects
 designed to detect and localize FRBs with arcsecond accuracy in real time
 are under construction/proposition, for example
 DSA-10~\cite{DSA-10}, DSA-2000~\cite{DSA-2000}, UTMOST-2D~\cite{UTMOST},
 MeerKAT~\cite{Booth:2009ex,Johnston:2020qxo,MeerTRAP}, and
 LOFAR~\cite{vanHaarlem:2013dsa}. It is expected that numerous
 FRBs with identified redshifts will be available in the future. Thus, it
 is reasonable to consider the simulated FRBs with known redshifts.

Let us briefly describe the steps to generate the simulated FRBs with
 known redshifts. At first, we should assign a random redshift $z_i$ to
 the $i$-th simulated FRB. To this end, the redshift distribution of FRBs
 should be assumed. In this work, we consider two types of redshift
 distributions for FRBs proposed in~\cite{Munoz:2016tmg}. The first
 one (termed ``\,Pzconst\,'') assumes that FRBs have a constant
 comoving number density, and the corresponding redshift distribution
 function reads~\cite{Munoz:2016tmg}
 \be{eq18}
 P_{\rm const}(z)\propto\frac{\chi^2(z)}{\left(1+z\right)H(z)}
 \,\exp\left(-\frac{d_L^{\,2}(z)}{2\,d_L^{\,2}(z_{\rm cut})}\right)\,,
 \ee
 where $\chi(z)=d_L(z)/(1+z)=c\int_0^z d\tilde{z}/H(\tilde{z})$ is the
 comoving distance. Gaussian cutoff at $z_{\rm cut}$ is introduced to
 represent an instrumental signal-to-noise threshold. The second one
 (termed ``\,PzSFH\,'') assumes that FRBs follow the star-formation
 history (SFH)~\cite{Caleb:2015uuk}, whose
 density is given by~\cite{Munoz:2016tmg}
 \be{eq19}
 \dot{\rho}_\ast(z)=\frac{\left(b_1+b_2 z\right)h}{\,1+
 \left(z/b_3\right)^{b_4\,}}\,,
 \ee
 with $b_1=0.0170$, $b_2=0.13$, $b_3=3.3$, $b_4=5.3$ and
 $h=0.7$~\cite{Cole:2000ea,Hopkins:2006bw,Munoz:2016tmg}. The corresponding
 redshift distribution function reads~\cite{Munoz:2016tmg}
 \be{eq20}
 P_{\rm SFH}(z)\propto
 \frac{\dot{\rho}_\ast(z)\,\chi^2(z)}{\,\left(1+z\right)H(z)}\,
 \exp\left(-\frac{d_L^{\,2}(z)}{2\,d_L^{\,2}(z_{\rm cut})}\right)\,.
 \ee
 In this work, we generate the simulated FRBs by using the simplest flat
 $\Lambda$CDM model as the fiducial cosmology, whose dimensionless Hubble
 parameter is given by
 \be{eq21}
 E(z)=H(z)/H_0=
 \left[\,\Omega_{m,\,0}(1+z)^3+(1-\Omega_{m,\,0})\,\right]^{1/2}\,,
 \ee
 where $\Omega_{m,\,0}$ is the present fractional density of matter
 (including cold dark matter and baryons). We adopt the most recent flat
 $\Lambda$CDM parameters from Planck 2018 CMB data~\cite{Aghanim:2018eyx},
 namely $H_0=67.36\;{\rm km/s/Mpc}$, $\Omega_{m,\,0}=0.3153$,
 and $\Omega_{b,\,0}=0.0493$. We adopt $z_{\rm cut}=0.5$
 following~\cite{Munoz:2016tmg}. In the left panel of Fig.~\ref{fig2},
 we show these two distributions as functions of redshift $z$. They are
 reasonable according to the crude rule of thumb $z\sim {\rm DM}/(1000
 \;\rm pc\hspace{0.24em}cm^{-3})<1.5$~\cite{Lorimer:2018rwi} for most of the
 observed FRBs to date having ${\rm DM_{obs}}<1500\;\rm pc\hspace{0.24em}
 cm^{-3}$~\cite{Petroff:2016tcr}. For the $i$-th simulated FRB, we can
 randomly assign a redshift $z_i$ to it from the redshift distributions
 Pzconst or PzSFH, which will be specified below.

The second step is to assign the corresponding ${\rm DM}_{{\rm IGM},\,i}$
 and its uncertainty $\sigma_{{\rm IGM},\,i}$ to this simulated FRB. To
 this end, we should preset several fiducial $f_{\rm IGM}(z)$ functions,
 for example $f_{\rm IGM}(z)={\rm const.}$ or
 $f_{\rm IGM}(z)=f_{\rm IGM,\,0}\left(1+\alpha\left(1-a\right)\right)=
 f_{\rm IGM,\,0}\left(1+\alpha\,z/(1+z)\right)$, which will be specified
 below. Then, we can calculate the mean $\langle{\rm DM_{IGM}}\rangle$ by
 using Eq.~(\ref{eq6}). As mentioned above, $\rm DM_{IGM}$ will deviate
 from $\langle{\rm DM_{IGM}}\rangle$ if the plasma density fluctuations
 are taken into account~\cite{McQuinn:2013tmc} (see also
 e.g.~\cite{Ioka:2003fr,Jaroszynski:2018vgh}). The uncertainty
 $\sigma_{\rm IGM}$ was studied in e.g.~\cite{McQuinn:2013tmc}, where
 three models for halo gas profile of the ionized baryons were used.
 Here, we consider the simplest one, namely the top hat model, and the
 corresponding $\sigma_{\rm IGM}$ was given by the green dots in the
 bottom panel of Fig.~1 of~\cite{McQuinn:2013tmc}. It is easy to fit
 these 27 green dots by using a very simple power law function
 \be{eq22}
 \sigma_{\rm IGM}(z)=173.8\, z^{0.4}\ {\rm pc\hspace{0.24em}cm^{-3}}\,.
 \ee
 In the right panel of Fig.~\ref{fig2}, we reproduce these 27 green dots
 from~\cite{McQuinn:2013tmc} and plot the power law $\sigma_{\rm IGM}(z)$
 given by Eq.~(\ref{eq22}). Clearly, they coincide with each other fairly
 well. For the $i$-th simulated FRB, we can randomly assign
 ${\rm DM}_{{\rm IGM},\,i}$ to it from a Gaussian distribution
 \be{eq23}
 {\rm DM}_{{\rm IGM},\,i}={\cal N}\left(\langle{\rm DM_{IGM}}
 \rangle (z_i),\,\sigma_{\rm IGM}(z_i)\right)\,,
 \ee
 while $\sigma_{{\rm IGM},\,i}=\sigma_{\rm IGM}(z_i)$. Obviously, we
 have $\rm DM_{IGM}=0$ at $z=0$ as expected by definition.

The third step is to assign ${\rm DM}_{{\rm HG},\,i}$ and its uncertainty
 $\sigma_{{\rm HG},\,i}$ to this simulated FRB. According to
 Eq.~(\ref{eq8}) and following e.g.~\cite{Yang:2016zbm,Gao:2014iva,
 Yang:2017bls,Zhou:2014yta,Qiang:2019zrs,Li:2019klc}, we have
 \be{eq24}
 {\rm DM}_{{\rm HG},\,i}={\rm DM}_{{\rm HG,\,loc},\,i}/(1+z_i)\,,\quad
 \quad\quad\sigma_{{\rm HG},\,i}=\sigma_{{\rm HG,\,loc},\,i}/(1+z_i)\,,
 \ee
 where ${\rm DM}_{{\rm HG,\,loc},\,i}$ can be randomly assigned from a
 Gaussian distribution with the mean $\langle{\rm DM_{HG,\,loc}}\rangle$
 and a fluctuation $\sigma_{\rm HG,\,loc}$~\cite{Yang:2016zbm,Gao:2014iva,
 Yang:2017bls,Zhou:2014yta,Qiang:2019zrs,Li:2019klc}, namely
 \be{eq25}
 {\rm DM}_{{\rm HG,\,loc},\,i}={\cal N}\left(\langle{\rm DM_{HG,\,loc}}
 \rangle,\,\sigma_{\rm HG,\,loc}\right)\,,\quad\quad {\rm and}
 \quad\quad\sigma_{{\rm HG,\,loc},\,i}=\sigma_{\rm HG,\,loc}\,.
 \ee
 In order to preset the fiducial values of
 $\langle{\rm DM_{HG,\,loc}}\rangle$ and $\sigma_{\rm HG,\,loc\,}$, it
 is helpful to examine the Milky Way. As is well known,
 $\rm DM_{MW}\lesssim 100\;\rm pc\hspace{0.24em}cm^{-3}$ at high Galactic
 latitude $|b|>10^\circ$, and its average dispersion is a few tens of
 $\rm pc\hspace{0.24em}cm^{-3}$~\cite{Taylor:1993my,Manchester:2004bp}
 (see also e.g.~\cite{Gao:2014iva,Zhou:2014yta}). Thus, it is reasonable
 to adopt the fiducial values $\langle{\rm DM_{HG,\,loc}}\rangle=100\;
 \rm pc\hspace{0.24em}cm^{-3}$ and $\sigma_{\rm HG,\,loc}=20\;\rm pc
 \hspace{0.24em}cm^{-3}$ following e.g.~\cite{Yang:2016zbm,Qiang:2019zrs}.

 \newpage  % used here just for a comfortable typesetting

%============================= Fig. 3 =================================

 \begin{center}
 \begin{figure}[tb]
 \centering
 \vspace{-10mm}  % used here just for a comfortable typesetting
 \includegraphics[width=0.85\textwidth]{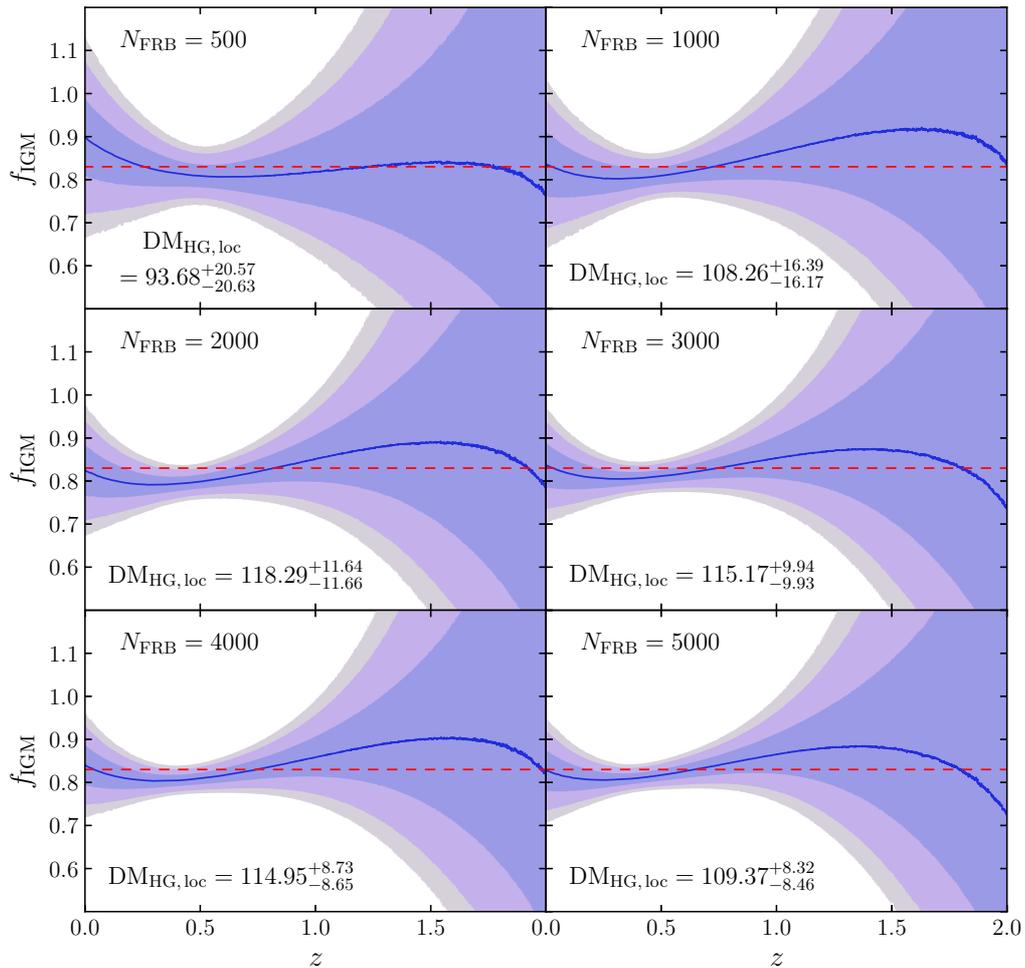}
 \caption{\label{fig3} $f_{\rm IGM}$ as functions of redshift
 $z$ reconstructed from various simulated FRB samples and the observational
 Pantheon SNIa sample. $N_{\rm FRB}$ is the number of simulated FRBs in
 each panel. The mean and $1\sigma$, $2\sigma$, $3\sigma$ uncertainties are
 indicated by the blue solid lines and the shaded regions, respectively.
 The reconstructed ${\rm DM_{HG,\,loc}}={\rm DM_{E\,}}|_{\,z=0}$ with
 $1\sigma$ uncertainties (in units of $\rm pc\hspace{0.24em}cm^{-3}$) are
 also presented in the corresponding panels. The preset $f_{\rm IGM}(z)$
 and redshift distribution used to generate these simulated FRB samples
 are $f_{\rm IGM}(z)=0.83$ and Pzconst, respectively. The red dashed lines
 indicate the preset $f_{\rm IGM}(z)$. See Sec.~\ref{sec3c} for details.}
 \end{figure}
 \end{center}

%======================================================================

%============================= Fig. 4 =================================

 \begin{center}
 \begin{figure}[tb]
 \centering
 \vspace{-10mm}  % used here just for a comfortable typesetting
 \includegraphics[width=0.85\textwidth]{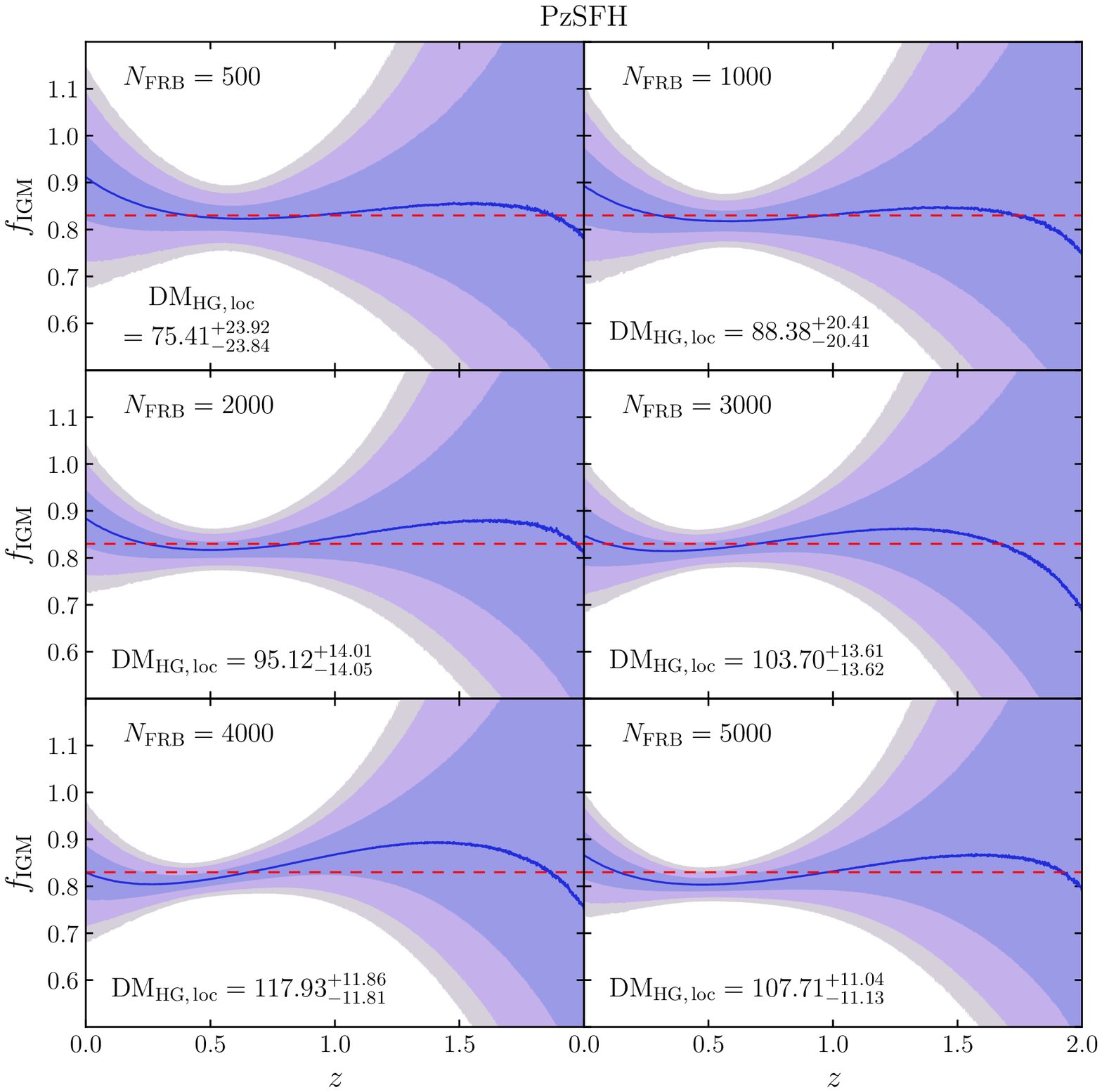}
 \caption{\label{fig4} The same as in Fig.~\ref{fig3}, but the preset
 $f_{\rm IGM}(z)$ and redshift distribution are $f_{\rm IGM}(z)=0.83$ and
 PzSFH, respectively. See Sec.~\ref{sec3c} for details.}
 \end{figure}
 \end{center}

%======================================================================

%============================= Fig. 5 =================================

 \begin{center}
 \begin{figure}[tb]
 \centering
 \vspace{-10mm}  % used here just for a comfortable typesetting
 \includegraphics[width=0.85\textwidth]{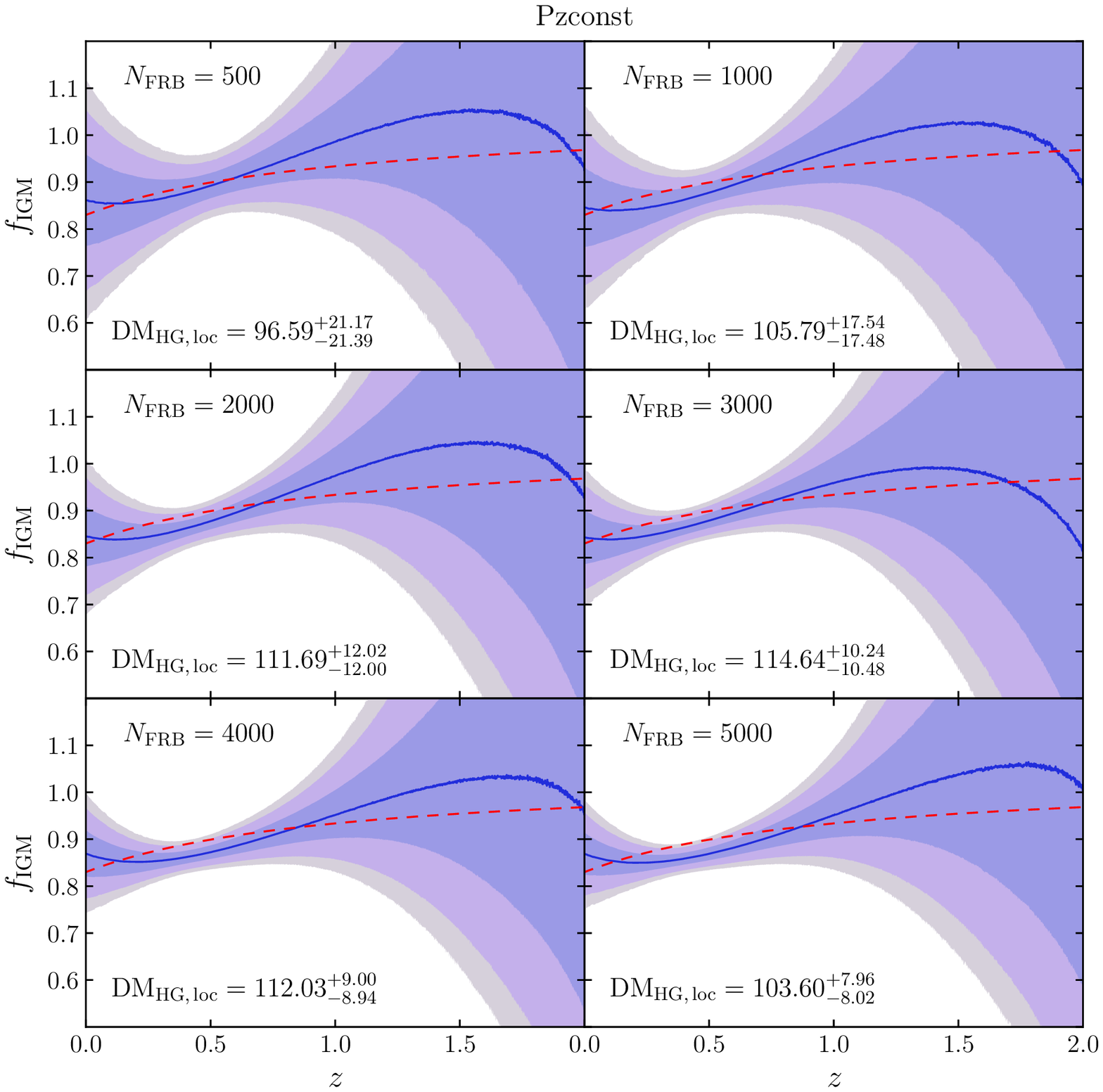}
 \caption{\label{fig5} The same as in Fig.~\ref{fig3}, but the preset
 $f_{\rm IGM}(z)$ and FRB redshift distribution are $f_{\rm IGM}(z)=
 0.83\,(1 + 0.25\,z/(1+z))$ and Pzconst, respectively. See
 Sec.~\ref{sec3c} for details.}
 \end{figure}
 \end{center}

%======================================================================

%============================= Fig. 6 =================================

 \begin{center}
 \begin{figure}[tb]
 \centering
 \vspace{-10mm}  % used here just for a comfortable typesetting
 \includegraphics[width=0.85\textwidth]{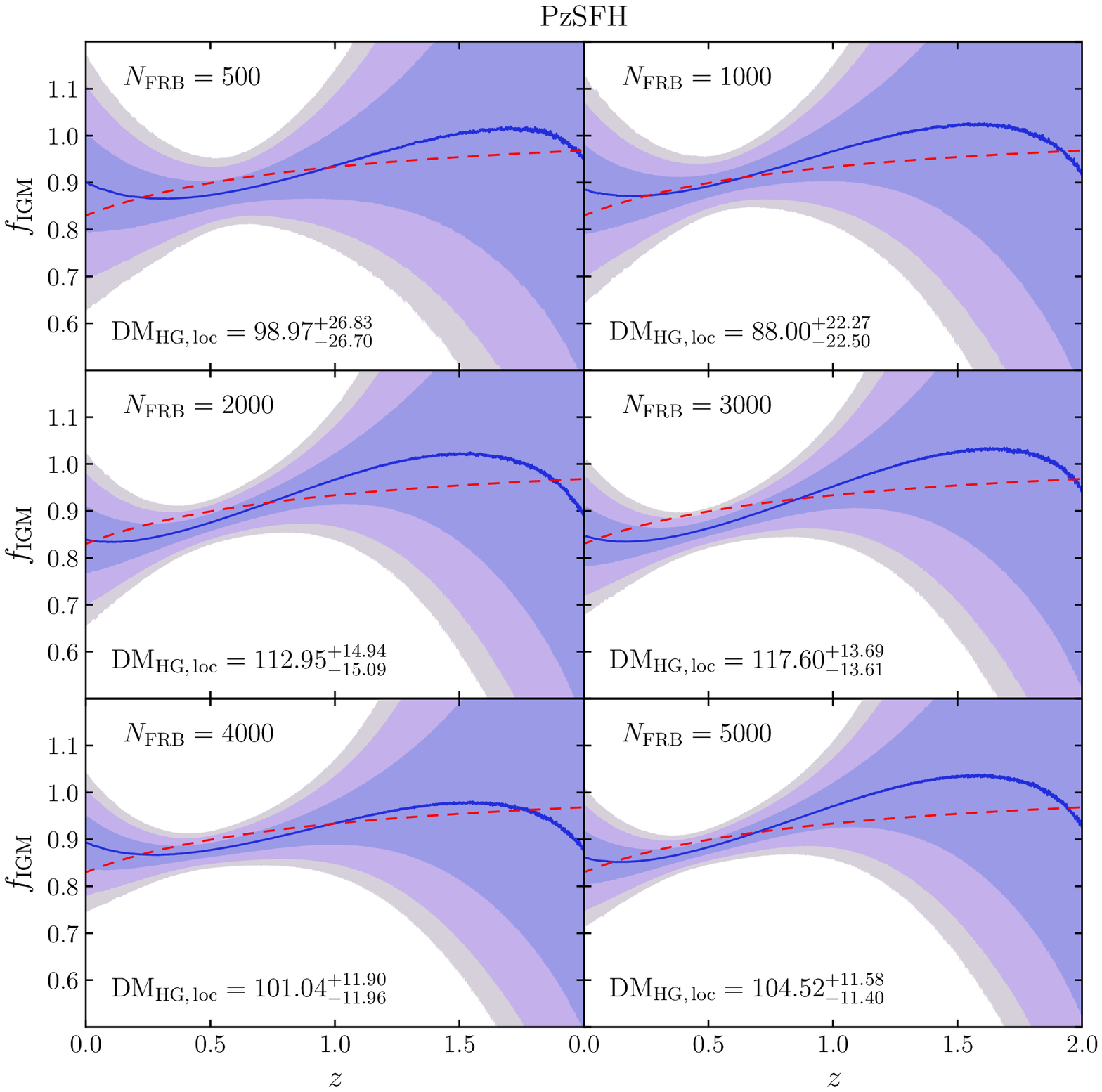}
 \caption{\label{fig6} The same as in Fig.~\ref{fig3}, but the preset
 $f_{\rm IGM}(z)$ and FRB redshift distribution are $f_{\rm IGM}(z)=
 0.83\,(1 + 0.25\,z/(1+z))$ and PzSFH, respectively. See
 Sec.~\ref{sec3c} for details.}
 \end{figure}
 \end{center}

%======================================================================

 \vspace{-30mm}  % used here just for a comfortable typesetting

Finally, the simulated $\rm DM_E$ data and its uncertainty
 for the $i$-th simulated FRB are given by
 \be{eq26}
 {\rm DM}_{{\rm E},\,i}={\rm DM}_{{\rm IGM},\,i}+{\rm DM}_{{\rm HG},\,i}\,,
 \quad\quad {\rm and}\quad\quad
 \sigma_{{\rm E},\,i}=(\sigma_{{\rm IGM},\,i}^2+
 \sigma_{{\rm HG},\,i}^2)^{1/2}\,.
 \ee
 One can repeat the above steps for $N_{\rm FRB}$ times to
 generate $N_{\rm FRB}$ simulated FRBs. The formatted data file
 for the simulated FRB sample contains $N_{\rm FRB}$ rows of $\{z_i$\,,
 ${\rm DM}_{{\rm E},\,i}$\,, $\sigma_{{\rm E},\,i}\}$. As mentioned at
 the beginning of this subsection, it is expected that numerous FRBs with
 identified redshifts will be available in the future. Thus, $N_{\rm FRB}$
 can be large, for example ${\cal O}(10^3)$ or even more.

%=========================== section 3.3 =================================

\subsection{Reconstructing the evolution of $f_{\rm IGM}(z)$}\label{sec3c}

We test our methodology by reconstructing the evolution of $f_{\rm IGM}(z)$
 with simulated FRB samples. We generate these samples following the
 procedures stated in Sec.~\ref{sec3b}, with the preset parameters, the
 specified $f_{\rm IGM}(z)$ and redshift distributions. Then, we reconstruct
 $f_{\rm IGM}(z)$ via Gaussian processes following the methodology given in
 Sec.~\ref{sec2a}, and also get $\rm DM_{HG,\,loc}$ from Eq.~(\ref{eq15}).
 Note that in Eq.~(\ref{eq14}) we use the reconstructed $E(z)$ from the
 observational Pantheon SNIa sample, as mentioned in Sec.~\ref{sec3a}.
 Finally, we check whether the reconstructed $f_{\rm IGM}(z)$
 and $\rm DM_{HG,\,loc}$ can be consistent with the ones
 used to generate the corresponding simulated FRB sample.

At first, we consider the simulated FRB samples with the preset $f_{\rm
 IGM}(z)=0.83$ (const.) and redshift distribution Pzconst, consisting of
 $N_{\rm FRB}=500$, $1000$, ..., $5000$ simulated FRBs, respectively. Note
 that the fiducial value of $f_{\rm IGM}$ of $0.83$ is chosen following
 e.g.~\cite{Yang:2016zbm,Yang:2017bls,Gao:2014iva,Qiang:2019zrs}. We
 present the reconstructed $f_{\rm IGM}(z)$ and ${\rm DM_{HG,\,loc}}=
 {\rm DM_E}|_{z=0}$ in Fig.~\ref{fig3}. Obviously, the uncertainties of
 the reconstructed $f_{\rm IGM}(z)$ are fairly large at high redshifts
 (especially at $z>1.2$). This is mainly due to the sparsity of simulated
 FRBs (and SNIa) data points at high redshifts (actually there are only a
 few data points at $z>1.2$ in the simulated samples, and FRBs at $z>1.5$
 are very rare (nb. the left panel of Fig.~\ref{fig2})). Thus, we mainly
 focus on the reconstructed $f_{\rm IGM}(z)$ at low redshift $z<1.2$. As
 expected, the uncertainties become smaller when the number of simulated
 FRBs $N_{\rm FRB}$ increases. From Fig.~\ref{fig3}, we see that the
 reconstructed $f_{\rm IGM}(z)$ and ${\rm DM_{HG,\,loc}}$ can be well
 consistent with the ones used to generate these simulated FRB
 samples, namely $f_{\rm IGM}(z)=0.83$ and ${\rm DM_{HG,\,loc}}=100\pm
 20\;\rm pc\hspace{0.24em}cm^{-3}$.

We turn to the simulated FRB samples with the preset $f_{\rm IGM}(z)=0.83$
 (const.) and redshift distribution PzSFH. The reconstructed
 $f_{\rm IGM}(z)$ and ${\rm DM_{HG,\,loc}}={\rm DM_E}|_{z=0}$ are given in
 Fig.~\ref{fig4}. It is easy to see that the difference between
 Figs.~\ref{fig4} and~\ref{fig3} is minor. For small $N_{\rm FRB}$, the
 means of reconstructed ${\rm DM_{HG,\,loc}}$ for the cases of PzSFH are
 slightly smaller than the ones for the cases of Pzconst, but they can be
 consistent with each other within $1\sigma$ uncertainties. The FRB
 redshift distributions (PzSFH and Pzconst) do not remarkably affect the
 reconstructions. In the cases of PzSFH, the reconstructed $f_{\rm IGM}(z)$
 and ${\rm DM_{HG,\,loc}}$ can also be well consistent with the ones used
 to generate these simulated FRB samples.

It is of interest to consider the cases of varying $f_{\rm IGM}(z)$.
 The simplest varying $f_{\rm IGM}(z)$ is given by a linear
 parameterization with respect to the scale factor $a$, namely
 $f_{\rm IGM}(z)=f_{\rm IGM,\,0}\,(1+\alpha\,(1-a))=
 f_{\rm IGM,\,0}\,(1+\alpha\,z/(1+z))$~\cite{Li:2019klc}. Actually this is
 reasonable, since a linear parameterization can be regarded as the Taylor
 series expansion up to the first order. Following~\cite{Li:2019klc},
 we preset the fiducial values $f_{\rm IGM,\,0}=0.83$ and $\alpha=0.25$.
 We generate the simulated FRB samples with this preset varying
 $f_{\rm IGM}(z)$ and redshift distribution Pzconst, and present the
 reconstructed $f_{\rm IGM}(z)$ and ${\rm DM_{HG,\,loc}}={\rm DM_E}|_{z=0}$
 in Fig.~\ref{fig5}. It is easy to see that the uncertainties
 of reconstructions become smaller when the number of simulated FRBs
 $N_{\rm FRB}$ increases. Clearly, the reconstructed $f_{\rm IGM}(z)$
 can successfully reproduce the rising tendency of the preset
 $f_{\rm IGM}(z)=0.83\,(1 + 0.25\,z/(1+z))$ as redshift $z$ increases.
 They are mutually consistent. Furthermore, the reconstructed
 ${\rm DM_{HG,\,loc}}$ can also be well consistent with
 the one used to generate these simulated FRB samples, namely
 ${\rm DM_{HG,\,loc}}=100\pm 20\;\rm pc\hspace{0.24em}cm^{-3}$.

Then, we turn to the cases of redshift distribution PzSFH, while the
 preset varying $f_{\rm IGM}(z)=0.83\,(1 + 0.25\,z/(1+z))$ is unchanged.
 We present the reconstructed $f_{\rm IGM}(z)$ and ${\rm DM_{HG,\,loc}}=
 {\rm DM_E}|_{z=0}$ in Fig.~\ref{fig6}. Once again, it is easy to see that
 the difference between Figs.~\ref{fig6} and~\ref{fig5} is minor. The FRB
 redshift distributions (PzSFH and Pzconst) do not remarkably affect the
 reconstructions. In the cases of PzSFH, the reconstructed $f_{\rm IGM}(z)$
 and ${\rm DM_{HG,\,loc}}$ can also be well consistent with the ones
 used to generate the simulated FRB samples.

%============================= section 4 ===================================

\section{Concluding remarks}\label{sec4}

FRBs are a promising new probe for astronomy and cosmology. Due to their
 extragalactic and cosmological origin, FRBs can be used to study IGM and
 the cosmic expansion. It is expected that numerous FRBs with identified
 redshifts will be available in the coming decade. $\rm DM_{IGM}$, the
 contribution from IGM to the observed DM of FRB, carries the
 information about the IGM and the cosmic expansion history. We can study
 the evolution of the universe by using FRBs with identified redshifts.
 In this work, we are interested in the fraction of baryon mass in IGM,
 $f_{\rm IGM}$, which is useful to study the cosmic expansion and the
 problem of missing baryons. We propose to reconstruct the evolution of
 $f_{\rm IGM}$ as a function of redshift $z$ with FRBs via a completely
 model-independent method, namely Gaussian processes. Since there is as
 yet no large sample of FRBs with identified redshifts, we use
 simulated FRBs instead. Through various simulations, we show that this
 methodology works well. The reconstructed $f_{\rm IGM}(z)$ and
 ${\rm DM_{HG,\,loc}}$ can be consistent with the ones used to generate
 the simulated FRB samples within $2\sigma$ and $1\sigma$ uncertainties,
 respectively, in the redshift range $0<z<1.2$.

As expected, the uncertainties become smaller as the number of simulated
 FRBs $N_{\rm FRB}$ increases. From Figs.~\ref{fig3}--\ref{fig6}, we find
 that the uncertainties become approximately stable for $N_{\rm FRB}\geq
 2000$, namely the improvement is not significant for more FRBs. The means
 of the reconstructed $f_{\rm IGM}(z)$ deviate from the preset ones by no
 more than approximately $8\%$ in the redshift range $0<z<1.2$ for all cases
 (it can be much better than $8\%$ for some particular cases). On the other
 hand, the uncertainties are fairly large for $N_{\rm FRB}\leq 1000$. Thus,
 we suggest that $1000\sim 2000$ FRBs are suitable for a fine
 model-independent reconstruction without assuming a particular function
 form or parameterization for $f_{\rm IGM}(z)$.

However, it might be many years before we have $1000\sim 2000$ FRBs with
 identified redshifts. Taking the projects such as CHIME, ASKAP, DSA,
 UTMOST-2D and MeerKAT into account, even 100 localized FRBs are some
 years away (we thank the referee for pointing out this issue). More
 powerful telescopes are desirable. We hope $1000\sim 2000$ FRBs with
 identified redshifts will be available in the coming decades.

The main source of the uncertainties is the large $\sigma_{\rm IGM}$.
 From Eq.~(\ref{eq22}) and the right panel of Fig.~\ref{fig2}, we have
 $\sigma_{\rm IGM}\gtrsim 120\;{\rm pc\hspace{0.24em}cm^{-3}}$ at redshifts
 $z>0.4$, and $\sigma_{\rm IGM}\gtrsim 150\;{\rm pc\hspace{0.24em}cm^{-3}}$
 at redshifts $z>0.7$. We hope that the statistical noise of
 FRBs (especially $\sigma_{\rm IGM}$) can be significantly reduced by the
 help of future developments. For example, for the lensed FRBs, one might
 infer the main contribution from the halo gas through gravitational
 lensing. With the reduced $\sigma_{\rm IGM}$, less FRBs (say, a few
 hundred) could be suitable for a fine model-independent reconstruction
 without assuming a particular function form or parameterization for
 $f_{\rm IGM}(z)$.

It is worth noting that in this work the observational Pantheon sample
 consisting of 1048 SNIa is used to reconstruct $E(z)=H(z)/H_0$, which
 is needed in Eq.~(\ref{eq14}). Actually, one can instead use some simulated
 samples consisting of a large number (say, $5000\sim 8000$) of SNIa with
 also much smaller uncertainties, which will be available in the future
 (especially in the era of WFIRST). In this case, it is natural to expect
 that the reconstructed $f_{\rm IGM}(z)$ might be much better than the ones
 obtained here. On the other hand, one can also use the observational or
 simulated $H(z)$ data, instead of SNIa, to reconstruct $E(z)=H(z)/H_0$. We
 anticipate that these will not change the main conclusions of this work.

Following e.g.~\cite{Yang:2016zbm,Yang:2017bls,Gao:2014iva,Qiang:2019zrs},
 in this work the fiducial value $f_{\rm IGM,\,0}=f_{\rm IGM}(z=0)=0.83$ is
 chosen, which is consistent with e.g.~\cite{Fukugita:1997bi,Shull:2011aa,
 Deng:2013aga}. However, there exist other values in the literature.
 For example, a smaller value $f_{\rm IGM}=0.6\pm 0.1$ was suggested
 in e.g.~\cite{Shull:2017eow}. Since we just use the fiducial value
 of $f_{\rm IGM,\,0}$ to generate the simulated FRBs, the exact value
 actually does not affect the discussions and the conclusions in this work.

In the present work, to generate the simulated FRBs, we have
 considered two types of the preset $f_{\rm IGM}(z)$, namely $f_{\rm IGM}(z)
 ={\rm const.}$ or a linear parameterization with respect to the scale
 factor $a$, i.e. $f_{\rm IGM}(z)=f_{\rm IGM,\,0}\,(1+\alpha\,(1-a))=
 f_{\rm IGM,\,0}\,(1+\alpha\,z/(1+z))$. Obviously, one can also consider
 other types of the preset $f_{\rm IGM}(z)$ instead, such as a linear
 parameterization with respect to the $e$-folding time $\ln a$, namely
 $f_{\rm IGM}(z)=f_{\rm IGM,\,0}\,(1-\alpha\ln a)=f_{\rm IGM,\,0}\,
 (1+\alpha\ln (1+z))$. Of course, $f_{\rm IGM}(z)$ as the Taylor series
 expansion up to higher order (say, 2nd order) with respect to the scale
 factor $a$ or the $e$-folding time $\ln a$ is also possible. Even the
 exotic types of the preset $f_{\rm IGM}(z)$ can also be considered, for
 instance an oscillating $f_{\rm IGM}(z)$. Note that these are just the
 preset $f_{\rm IGM}(z)$ used to generate the simulated FRBs. Instead, the
 real $f_{\rm IGM}(z)$ of the universe will be reconstructed or determined
 by using the real FRBs with identified redshifts in the future. In doing
 this, we need not assume any specific function form or
 parameterization for $f_{\rm IGM}(z)$, because Gaussian processes are
 completely model-independent.

%============================= acknowledgements ==================================

\section*{ACKNOWLEDGEMENTS}

We heartily thank the anonymous referee for all the very expert
 and useful comments and suggestions, which have significantly helped us
 to improve this work. We are grateful to Zhao-Yu~Yin, Hua-Kai~Deng,
 Zhong-Xi~Yu and Shu-Ling~Li for kind help and useful discussions. This
 work was supported in part by NSFC under Grants No.~11975046
 and No.~11575022.

\renewcommand{\baselinestretch}{1.01}

%============================= references =================================

\end{document}